\documentclass[10pt,journal,cspaper,compsoc]{IEEEtran}

\usepackage{amssymb,url}
\usepackage{graphicx,epsfig,color,wrapfig}
\usepackage{listings}
\title{Chernoff information of exponential families}

\author{Frank Nielsen,~\IEEEmembership{Senior Member,~IEEE,}
\IEEEcompsocitemizethanks{\IEEEcompsocthanksitem F. Nielsen is with the Sony Computer Science Laboratories (Tokyo, Japan) and \'Ecole Polytechnique (Palaiseau, France).\protect\\
E-mail: nielsen@lix.polytechnique.fr }
}

\date{\today}

\def\KL{\mathrm{KL}}
\def\dx{\mathrm{d}x}
\def\Pr{\mathrm{Pr}}
\def\Error{\mathrm{Error}}
\def\innerproduct#1#2{ \langle {#1},{#2} \rangle }
\def\Innerproduct#1#2{ \left\langle {#1},{#2} \right\rangle }

\def\tr{\mathrm{tr}}

\def\D{\mathfrak{D}}

\begin{document}

\IEEEcompsoctitleabstractindextext{%
\begin{abstract}
Chernoff information upper bounds   the probability of error of the optimal Bayesian decision rule for $2$-class classification problems. 
However,  it turns out that  in practice the Chernoff bound is hard to calculate or even approximate.
In statistics, many usual  distributions, such as Gaussians, Poissons or frequency histograms called multinomials, can be handled in the unified framework of exponential families.
In this note, we prove that the Chernoff information for members of the same exponential family can be either derived  analytically in closed form, or  efficiently approximated using a simple geodesic bisection optimization technique based on an exact geometric characterization of the ``Chernoff point'' on the underlying  statistical   manifold. 
\end{abstract}

}

\maketitle

\IEEEdisplaynotcompsoctitleabstractindextext
\IEEEpeerreviewmaketitle

\section{Introduction}

\IEEEPARstart{C}{onsider}
 the following statistical decision problem of {\it classifying} a random observation $x$ as one of two possible classes: 
$C_1$ and $C_2$ (say, detect target signal from noise signal).
Let $w_1=\Pr(C_1)>0$ and $w_2=\Pr(C_2)=1-w_1>0$ denote the {\it a priori} class probabilities, and let $p_1(x)=\Pr(x|C_1)$ and $p_2(x)=\Pr(x|C_2)$ denote the {\it class-conditional} probabilities, so that we have $p(x)=w_1p_1(x)+w_2p_2(x)$.  
{\it Bayes decision rule} classifies $x$ as $C_1$ if $\Pr(C_1|x)>\Pr(C_2|x)$, and as $C_2$ otherwise.
Using Bayes rule\footnote{Bayes rule states that the joint probability of two events equals the product of the probability of one event times the conditional probability of the second event given the first one. That is, in mathematical terms 
$\Pr(x\wedge \theta)=\Pr(x)\Pr(\theta|x)=\Pr(\theta)\Pr(x|\theta)$, so that we have
$\Pr(\theta|x)=\Pr(\theta)\Pr(x|\theta) / \Pr(x)$.}, 
we have $\Pr(C_i|x)=\frac{\Pr(C_i)\Pr(x|C_i)}{\Pr(x)}= \frac{w_i p_i(x)}{p(x)}$ for $i\in\{1,2\}$.
Thus Bayes decision rule assigns $x$ to class $C_1$ if and only if $w_1 p_1(x) > w_2 p_2(x)$, and to $C_2$ otherwise.
Let $L(x)=\frac{\Pr(x|C_1)}{\Pr(x|C_2)}$ denote the {\it likelihood ratio}.
In decision theory~\cite{ct-1991}, Neyman and Pearson proved that the optimum decision {\it test} has necessarily to be of the form $L(x)\geq t$ to accept hypothesis $C_1$, where $t$ is a threshold value. 

The probability of error $E=\Pr(\Error)$ of {\it any} decision rule $\D$ is $E=\int p(x)\Pr(\Error|x) \dx$, where 

$$
\Pr(\Error|x) = \left\{
\begin{array}{ll}
\Pr(C_1|x) & \mbox{if $\D$ wrongly decided $C_2$},\\
\Pr(C_2|x) & \mbox{if $\D$ wrongly decided $C_1$}.
\end{array}
\right.
$$

Thus Bayes decision rule minimizes {\it by principle} the average {\em probability of error}:

\begin{eqnarray}
E^* &=& \int \Pr(\Error | x) p(x) \dx,\\
& =& \int \min(\Pr(C_1|x), \Pr(C_2|x)) p(x) \dx.
\end{eqnarray}

The Bayesian rule is also called the maximum a-posteriori (MAP) decision rule.
Bayes error constitutes therefore the reference benchmark since no other decision rule can beat its classification performance.

Bounding tightly the Bayes error is  thus crucial in  hypothesis testing.
Chernoff derived a notion of information\footnote{In information theory, there exists several notions of information such as Fisher information in Statistics or Shannon information in Coding theory. Those various definitions gained momentum by asking questions like "How hard is it to estimate/discriminate distributions?" (Fisher) or "How hard is it to compress data?" (Shannon). Those "how hard..." questions were answered by proving lower bounds (Cram\'er-Rao for Fisher, and Entropy for Shannon).
Similarly, Chernoff information answers the "How hard is it to classify (empirical) data?" by providing a tight lower bound: the (Chernoff) (classification) information.}  
from this hypothesis task (see Section~7 of~\cite{Chernoff-1952}).
To upper bound Bayes error,  one replaces the minimum function by a smooth power function:
Namely, for $a,b> 0$, we have

\begin{equation}
\min (a,b)\leq a^\alpha b^{1-\alpha},\forall\alpha\in(0,1).
\end{equation}
Thus we get the following Chernoff bound: 

\begin{eqnarray}
E^* &= &\int \min(\Pr(C_1|x), \Pr(C_2|x)) p(x) \dx \\
& \leq & w_1^{\alpha} w_2^{1-\alpha}  \int p_1^{\alpha}(x) p_2^{1-\alpha}(x) \dx
\end{eqnarray}

Since the inequality holds for any $\alpha\in (0,1)$, we  upper bound the minimum error $E^*$ as follows

$$
E^* \leq w_1^{\alpha} w_2^{1-\alpha}  c_\alpha(p_1 : p_2),
$$
where $c_\alpha(p_1:p_2)=    \int p_1^{\alpha}(x)p_2^{1-\alpha}(x) \dx$ is called the Chernoff $\alpha$-coefficient.
We use the ":" delimiter to emphasize the fact that this statistical measure is usually not symmetric: $c_{\alpha}(p_1 : p_2) \not = c_{\alpha}(p_2 : p_1)$, although we have $c_{\alpha}(p_2 : p_1)= c_{1-\alpha}(p_1 : p_2)$.
For $\alpha=\frac{1}{2}$, we obtain the symmetric Bhattacharrya coefficient~\cite{Bhatta1943} $b(p_1 : p_2)=c_{\frac{1}{2}}(p_1 : p_2)=\int \sqrt{p_1(x)p_2(x)}\dx=b(p_2,p_1)$.
The optimal Chernoff $\alpha$-coefficient is found by choosing the {\it best} exponent for upper bounding Bayes error~\cite{ct-1991}:

\begin{equation}
c^*(p_1 : p_2) = c_{\alpha^*}(p_1 : p_2)=   \min_{\alpha\in(0,1)} \int p_1^{\alpha}(x)p_2^{1-\alpha}(x) \dx.
\end{equation}

Since the Chernoff coefficient is a {\it measure of similarity} (with $0<c_\alpha(p_1,p_2)\leq 1$) relating to the overlapping of the   densities $p_1$ and $p_2$, it follows that we can derive thereof a statistical distance measure, called the {\it Chernoff information} (or Chernoff divergence) as

\begin{eqnarray}
C^*(p_1 : p_2) & =&  C_{\alpha^*}(p_1 : p_2) \\
& = &   -\log \min_{\alpha\in (0,1)} \int p_1^{\alpha}(x)p_2^{1-\alpha}(x) \dx \geq 0.\nonumber\\
 &= & \max_{\alpha\in (0,1)} -\log  \int p_1^{\alpha}(x)p_2^{1-\alpha}(x) \dx
\end{eqnarray}

In the remainder, we call Chernoff divergence (or Chernoff information) the measure $C^*(\cdot : \cdot)$, and Chernoff $\alpha$-divergence (of the first type) the functional $C_\alpha(p:q)$ (for $\alpha\in(0,1)$).
Chernoff information yields the best achievable exponent for a Bayesian probability of error~\cite{ct-1991}: 
\begin{equation}
E^*\leq w_1^{\alpha^*}w_2^{1-\alpha^*}e^{-C^*(p_1:p_2)}.
\end{equation}

From the Chernoff $\alpha$-coefficient measure of similarity, we can derive a second type of Chernoff $\alpha$-divergences~\cite{alphadiv-2010}  defined by $C'_\alpha(p:q)=\frac{1}{\alpha(1-\alpha)}(1-c_\alpha(p:q))$.
Those second type Chernoff $\alpha$-divergences are related to Amari $\alpha$-divergences~\cite{alphabetagamma-2010} by a linear mapping~\cite{alphadiv-2010} on the exponent $\alpha$, and to R\'enyi and Tsallis relative entropies (see Section~\ref{sec:ChernoffCoeff}).
In the remainder, Chernoff $\alpha$-divergences refer to the first-type divergence.


In practice, we do {\it not} have  statistical knowledge of the prior distributions of classes nor of the class-conditional distributions.
But we are rather given a training set of correctly labeled class points.
In that case, a simple decision rule, called the {\it nearest neighbor rule}\footnote{The nearest neighbor rule postulates that things that ``look alike must be alike.'' See~\cite{CoverHart-1967}.}, consists for an observation $x$, to label it according to the label of its nearest neighbor (ground-truth).
It can be shown that the probability error of this simple scheme is upper bounded by {\it twice} the optimal Bayes error~\cite{CoverHart-1967,Stone-1977}.
Thus half of the Chernoff information is contained somehow in the nearest neighbor knowledge, a key component of machine learning algorithms.
(It is traditional to improve this classification by taking a majority vote over the $k$ nearest neighbors.)

Chernoff information has appeared in many applications ranging from sensor networks~\cite{ChernoffInf-2003} to visual computing tasks such as image segmentation~\cite{ChernoffInf-2010},   image registration~\cite{stitching-1982}, face recognition~\cite{chernoffface-2005}, feature detector~\cite{ChernoffInf-2007}, and  edge segmentation~\cite{ChernoffInf-Vision2003}, just to name a few.

The paper is organized as follows: 
Section~\ref{sec:statdist} introduces the functional parametric Bregman and Jensen class of statistical distances. 
Section~\ref{sec:expfam} concisely describes the exponential families in statistics.
Section~\ref{sec:ChernoffCoeff} proves that the Chernoff $\alpha$-divergences of two members of the {\it same} exponential family class is equivalent to a skew Jensen divergence evaluated at the corresponding distribution parameters.  
In section~\ref{sec:JensenBregman}, we show that the optimal Chernoff coefficient obtained by minimizing skew Jensen divergences yields an equivalent Bregman divergence, which can be derived from a simple optimality criterion. 
It follows a closed-form formula for the Chernoff information on single-parametric exponential families in Section~\ref{sec:order1}.
We extend the optimality criterion to the multi-parametric case in  Section~\ref{sec:multiorder}.
Section~\ref{sec:chernoffpoint} characterizes geometrically the optimal solution by introducing concepts of information geometry.
Section~\ref{sec:primaldual} designs a simple yet efficient geodesic bisection search algorithm for approximating the multi-parametric case.
Finally, section~\ref{sec:Conclusion} concludes the paper.

\section{Statistical divergences }\label{sec:statdist}

Given two probability distributions with respective densities $p$ and $q$, a divergence $D(p : q)$ measures the distance between those distributions.
The classical divergence in information theory~\cite{ct-1991} is the {\it Kullback-Leibler divergence}, also called {\it relative entropy}:

\begin{equation}\label{eq:kl}
\KL(p : q)=\int p(x)\log \frac{p(x)}{q(x)} \dx
\end{equation}
(For probability mass functions, the integral is replaced by a discrete sum.)
This divergence is oriented (ie. $\KL(p : q)\not = \KL(q : p)$) and does not satisfy the triangle inequality of metrics.
It turns out that the Kullback-Leibler divergence belongs to a wider class of divergences called Bregman divergences.
A Bregman divergence is obtained for a strictly convex and differentiable generator $F$ as:

\begin{eqnarray}\label{eq:bd}
\lefteqn{B_F(p : q)=}\\
&&\int \left( F(p(x))-F(q(x))-(p(x)-q(x))F'(q(x))  \right) \dx\nonumber
\end{eqnarray}

The Kullback-Leibler divergence is obtained for the generator $F(x)=x\log x$, the negative Shannon entropy (also called Shannon information).
This functional parametric class of Bregman divergences $B_F$ can further be interpreted as  {\it limit cases} of skew Jensen divergences.
A skew Jensen divergence (Jensen $\alpha$-divergences, or $\alpha$-Jensen divergences)  is defined for a strictly convex generator $F$ as

\begin{eqnarray}
J_{F}^{(\alpha)}(p : q) &=& \int \left( \alpha F(p(x))+ (1-\alpha)F(q(x))- \right.\nonumber \\
 && \left. F(\alpha p(x)+(1-\alpha) q(x)) \right)  \dx \geq 0,  \nonumber\\
 && \forall\alpha\in(0,1) 
\end{eqnarray}

%

Note that $J_{F}^{(\alpha)}(p : q)=J_{F}^{(1-\alpha)}(q:p)$, and that $F$ is defined up to affine terms.
For $\alpha\to \{0,1\}$, the Jensen divergence tend to zero, and loose its power of discrimination.
However, interestingly, we have $\lim_{\alpha\to 1} J_{F}^{(\alpha)}(p : q)= \frac{1}{1-\alpha} B_F(p:q) $ and $\lim_{\alpha\to 0} J_{F}^{(\alpha)}(p : q)= \frac{1}{\alpha} B_F(q:p)$, as proved in~\cite{entdivmean-1995,2010-brbhat}. That is, Jensen $\alpha$-divergences tend asymptotically to (scaled) Bregman divergences.

The Kullback-Leibler divergence also belongs to the class of Csisz\'ar $F$-divergences (with $F(x)=x\log x$), defined for a convex function $F$ with $F(1)=0$:

\begin{equation}
I_F(p:q) = \int_x  F\left(\frac{p(x)}{q(x)} \right)  q(x) \dx.
\end{equation}
Amari's $\alpha$-divergences are the canonical divergences in $\alpha$-flat spaces in information geometry~\cite{informationgeometry-2000} defined by

\begin{equation}
A_\alpha(p: q) = \left \{
\begin{array}{lr}
 \frac{4}{1-\alpha^2} ( 1 - c_{\frac{1-\alpha}{2}}(p:q)), &  \alpha\not = \pm 1,\\
 \int p(x)\log \frac{p(x)}{q(x)} \dx= \KL(p,q) , & \alpha=-1,\\
 \int q(x)\log \frac{q(x)}{p(x)} \dx= \KL(q,p) , & \alpha=1,\\
 \end{array}
 \right .
\end{equation}
Those Amari $\alpha$-divergences (related to Chernoff $\alpha$-coefficients, and Chernoff $\alpha$-divergences of the second type by a linear mapping of the exponent~\cite{alphadiv-2010}) are $F$-divergences for the generator $F_\alpha(x)=\frac{4}{1-\alpha^2} (1-x^{\frac{1+\alpha}{2}})$, $\alpha\not\in\{-1,1\}$.

Next, we introduce a versatile class of probability densities in statistics for which $\alpha$-Jensen divergences (and hence Bregman divergences) admit closed-form formula.

\section{Exponential families}\label{sec:expfam}

A generic class of statistical distributions encapsulating many usual distributions (Bernoulli, Poisson, Gaussian, multinomials, Beta, Gamma, Dirichlet, etc.) are the exponential families.
We recall their elementary definition here, and refer the reader to~\cite{ef-flashcards-2009} for a more detailed overview.
An {\it exponential family} $E_F$ is a parametric set of probability distributions  admitting the following canonical decomposition of their densities:
\begin{equation}
	p(x; \theta) = \exp \left( \langle t(x), \theta \rangle - F(\theta) + k(x) \right)
\end{equation}
where
 $t(x)$ is the sufficient statistic,
 $\theta\in\Theta$ are the natural parameters belonging to an open convex natural space $\Theta$,
$\langle .,. \rangle$ is the inner product (i.e., $\langle x,y \rangle = x^T y$ for column vectors),
$F(\cdot)$ is the log-normalizer (a $C^{\infty}$ convex function),
and $k(x)$ the carrier measure.


For example, Poisson distributions $\Pr(x=k; \lambda)=\frac{\lambda^k e^{-\lambda}}{k!}$, 
for $k\in\mathbb{N}$ form an exponential family $E_F=\{p_F(x;\theta)\ |\ \theta\in\Theta\}$, with
 $t(x)=x$   the sufficient statistic, $\theta=\log\lambda$  the natural parameters,
$F(\theta)=\exp\theta$   the log-normalizer,
and $k(x)=-\log x!$ is the carrier measure. 

Since we often deal with applications using multivariate normals, we also report the canonical decomposition for the multivariate Gaussian family. 
We rewrite the Gaussian density of mean $\mu$ and variance-covariance matrix $\Sigma$:
$$
p(x;\mu,\Sigma) = \frac{1}{2\pi\sqrt{\det\Sigma}} \exp \left( - \frac{(x-\mu)^T\Sigma^{-1} (x-\mu))}{2} \right)
$$
in the canonical form with 
 $\theta=(\Sigma^{-1}\mu, \frac{1}{2}\Sigma^{-1})\in \Theta=\mathbb{R}^d\times\mathbb{K}_{d\times d}$ ($\mathbb{K}_{d\times d}$ denotes the cone of positive definite matrices),
$F(\theta)=\frac{1}{4}\tr(\theta_2^{-1} \theta_1\theta_1^T)-\frac{1}{2}\log\det \theta_2+\frac{d}{2}\log\pi$ the log-normalizer,
$t(x)=(x,-x^T x)$ the sufficient statistics, and $k(x)=0$ the carrier measure.
In that case, the inner product $\innerproduct{\cdot}{\cdot}$ is composite, and calculated as the sum of a vector dot product with a matrix trace product: $\innerproduct{\theta}{\theta'}=\theta_1^T\theta_1'+\tr(\theta_2^T \theta_2')$, where $\theta=[\theta_1\ \theta_2]^T$ and $\theta'=[\theta_1'\ \theta_2']^T$.
%
%

The {\it order} of an exponential family denotes the dimension of its parameter space.
For example, Poisson family is of order $1$, univariate Gaussians of order $2$, and $d$-dimensional multivariate Gaussians of order $\frac{d(d+3)}{2}$.
Exponential families brings mathematical convenience to easily solve tasks, like finding the maximum likelihood estimators~\cite{ef-flashcards-2009}.
It can be shown that the Kullback-Leibler divergence of members of the same exponential family is equivalent to a Bregman divergence on the natural parameters~\cite{bvd-2010}, thus bypassing the fastidious integral computation of Eq.~\ref{eq:kl}, and yielding a closed-form formula (following Eq.~\ref{eq:bd}):

\begin{equation}
\KL(p_F(x;\theta_p) : p_F(x;\theta_q) ) = B_F(\theta_q : \theta_p).
\end{equation}
Note that on the left hand side, the Kullback-Leibler is a distance acting on distributions, while on the right hand side, the Bregman divergence is a distance acting on corresponding swapper parameters. 

Exponential families play a crucial role in statistics as they also bring mathematical convenience for generalizing results.
For example, the log-likelihood ratio test for members of the same exponential family writes down as:

\begin{equation}
\log \frac{e^{\innerproduct{t(x)}{\theta_1}-F(\theta_1)+k(x)}}{ e^{\innerproduct{t(x)}{\theta_2}-F(\theta_2)+k(x)}} \geq \log \frac{w_2}{w_1}
\end{equation}

Thus the decision border is a {\it linear bisector} in the sufficient statistics $t(x)$:
\begin{equation}
\innerproduct{t(x)}{\theta_1-\theta_2}-F(\theta_1)+F(\theta_2)=\log \frac{w_2}{w_1}.
\end{equation}

\section{Chernoff coefficients of exponential families}\label{sec:ChernoffCoeff}

Let us prove that the Chernoff $\alpha$-divergence of members of the {\it same}  exponential families is equivalent to
 a $\alpha$-Jensen divergence defined for the log-normalizer generator, and evaluated at the corresponding natural parameters.
Without loss of generality, let us  consider the reduced canonical form of exponential families 
$p_F(x;\theta)=\exp(\innerproduct{x}{\theta}-F(\theta))$ (assuming $t(x)=x$ and $k(x)=0$).
Consider the Chernoff $\alpha$-coefficient of similarity of two distributions $p$ and $q$ belonging to the {\it same} exponential family $E_F$:

{\small
\begin{equation}
c_\alpha(p : q) =    \int p^{\alpha}(x) q^{1-\alpha}(x) \dx = \int p_F^{(\alpha)}(x;\theta_p) p_F^{1-\alpha}(x;\theta_q) \dx \nonumber
\end{equation}
\begin{eqnarray*}
 = && \int \exp (\alpha(\innerproduct{x}{\theta_p}-F(\theta_p)))  \exp ((1-\alpha)(\innerproduct{x}{\theta_q}-F(\theta_q))) \dx\\
 = && \int \exp \left( \Innerproduct{x}{\alpha\theta_p+(1-\alpha)\theta_q}-(\alpha F(\theta_p)+(1-\alpha)F(\theta_q) \right)\dx\\
 =  && \exp -(\alpha F(\theta_p)+(1-\alpha)F(\theta_q)) \int \exp ( \innerproduct{x}{\alpha\theta_p+(1-\alpha)\theta_q}
\\ && -F(\alpha\theta_p+(1-\alpha)\theta_q)+F(\alpha\theta_p+(1-\alpha)\theta_q) ) \dx\\
=  &&  \exp \left( F(\alpha\theta_p+(1-\alpha)\theta_q) - (\alpha F(\theta_p)+(1-\alpha)F(\theta_q) \right) \times \\ && \int \exp \innerproduct{x}{\alpha\theta_p+(1-\alpha)\theta_q}  -F(\alpha\theta_p+(1-\alpha)\theta_q)\dx \\
 = && \exp ( F(\alpha\theta_p+(1-\alpha)\theta_q) - \\ && (\alpha F(\theta_p)+(1-\alpha)F(\theta_q)) \times \int p_F(x;\alpha\theta_p+(1-\alpha)\theta_q)\dx\\
 = &&  \exp (   -J_F^{(\alpha)}(\theta_p : \theta_q)   ) \geq 0.
\end{eqnarray*}
}

It follows that the Chernoff $\alpha$-divergence (of the first type) is given by

\begin{eqnarray*}
C_\alpha(p : q)  & = &  -\log c_\alpha(p,q)  = J_{F}^{(\alpha)}(\theta_p : \theta_q),\\
c_\alpha(p : q)  & = & e^{-C_\alpha(p:q)} =e^{-J_{F}^{(\alpha)}(\theta_p : \theta_q)}.
\end{eqnarray*}

That is, the Chernoff $\alpha$-divergence on members of the same exponential family is equivalent to a  Jensen $\alpha$-divergence on the corresponding natural parameters.
For multivariate normals, we thus retrieve easily the following Chernoff $\alpha$-divergence between $p\sim N(\mu_1,\Sigma_1)$ and $q\sim N(\mu_2,\Sigma_2)$:

\begin{eqnarray}\label{eq:alphadivmvn}
&&C_\alpha(p,q) =  \frac{1}{2}\log \frac{|\alpha\Sigma_1+(1-\alpha)\Sigma_2|}{|\Sigma_1|^{\alpha} |\Sigma_2|^{1-\alpha}}  +\nonumber\\
&&\frac{\alpha(1-\alpha)}{2} (\mu_1-\mu_2)^T (\alpha\Sigma_1+(1-\alpha)\Sigma_2) (\mu_1-\mu_2).\nonumber\\
\ 
\end{eqnarray}
For $\alpha=\frac{1}{2}$, we find the Bhattacharyya distance~\cite{Bhatta1943,kailath-1967} between multivariate Gaussians.

Note that since Chernoff $\alpha$-divergences are related to R\'enyi $\alpha$-divergences 
\begin{equation}
R_{\alpha}(p : q) =  \frac{1}{\alpha-1} \log \int_x p(x)^\alpha q^{1-\alpha}(x) \dx,
\end{equation}
 built on 
 R\'enyi  entropy 
 \begin{equation}
 H_R^\alpha(p)  =  \frac{1}{1-\alpha}\log (\int_x p^\alpha(x)\dx -1),
 \end{equation}
  (and hence by a monotonic mapping\footnote{The Tsallis entropy $H_T^\alpha(p)=\frac{1}{\alpha-1}(1-\int p(x)^\alpha \dx)$
is obtained from the R\'enyi entropy (and vice-versa) via the mappings:
$H_T^\alpha(p)  =  \frac{1}{1-\alpha} (e^{(1-\alpha)H_R^\alpha(p)}-1)$ and
$H_R^\alpha(p) =  \frac{1}{1-\alpha} \log (1+(1-\alpha) H_T^\alpha(p))$.} to Tsallis divergences), closed form formulas for members of the same exponential family  follow:

\begin{eqnarray}
R_{\alpha}(p : q) & = &  \frac{1}{1-\alpha} C_{\alpha}(p:q),\\
R_{\alpha}(p_F(x;\theta_p): p_F(x;\theta_q)) &= &\frac{1}{1-\alpha} J_F^{(\alpha)}(\theta_p:\theta_q)\label{eq:rcfef}
\end{eqnarray}
(Note that   $R_{\frac{1}{2}}(p:q)$ is twice the Bhattacharyya coefficient: $R_{\frac{1}{2}}(p:q)=2C_{\frac{1}{2}}(p:q)$.)
 For example, the R\'enyi divergence on members $p\sim N(\mu_p,\Sigma_p)$ and $q\sim N(\mu_q,\Sigma_q)$ of the normal exponential family  is obtained in closed form solution using Eq.~\ref{eq:rcfef}:

\begin{eqnarray}
R_\alpha(p,q)=\frac{1}{2\alpha} (\mu_p-\mu_q)^T ((1-\alpha)\Sigma_p+\alpha\Sigma_q)^{-1}  (\mu_p-\mu_q)
+\nonumber\\ 
\frac{1}{1-\alpha}  \log \frac{\det((1-\alpha) \Sigma_p+\alpha\Sigma_q) }{\det(\Sigma_p^{1-\alpha})\det(\Sigma_q^{\alpha})}.
\end{eqnarray}

Similarly, for the Tsallis relative entropy, we have:
\begin{eqnarray}
T_{\alpha}(p:q) & = & \frac{1}{1-\alpha} (1-c_{\alpha}(q:p)),\\
T_{\alpha}(p_F(x;\theta_p): p_F(x;\theta_q)) & = &  \frac{(1-e^{-J_F^{(\alpha)}(q:p)})}{1-\alpha}
\nonumber\\
\ 
\end{eqnarray} 
Note that $\lim_{\alpha\to 1} R_\alpha(p:q)= \lim_{\alpha\to 1} T_\alpha(p:q) = \KL(p:q) = B_F(\theta_q:\theta_p)$, as expected.

So far, particular cases of exponential families have been considered for computing the Chernoff $\alpha$-divergences (but not Chernoff divergence).
For example, Rauber et al.~\cite{chernoffDirichlet-2008} investigated  statistical distances for Dirichlet and Beta distributions (both belonging to the exponential families). 
The density of a Dirichlet distribution parameterized by a $d$-dimensional vector $p=(p_1, ..., p_d)$ is

$$
\Pr(X=x;p)=\frac{\Gamma(\sum_{i=1}^d p_i)}{\prod_{i=1}^d \Gamma(p_i)} \prod_{i=1}^d x_i^{p_i-1},
$$
with $\Gamma(t)=\int_0^\infty z^{t-1} e^{-z} \mathrm{d}z$ the gamma function generalizing the factorial $\Gamma(n-1)=n!$.
Beta distributions are particular cases of Dirichlet distributions, obtained for $d=2$.
Rauber et al.~\cite{chernoffDirichlet-2008} report the following closed-form formula for the Chernoff $\alpha$-divergences:

\begin{eqnarray*}
C_\alpha(p:q)&=& \log \Gamma(\sum_{i=1}^d (\alpha p_{i} - (1-\lambda) q_i))\\
&&+\alpha \sum_{i=1}^d  \log \Gamma(p_i)  +
(1-\alpha) \sum_{i=1}^d  \log\Gamma(q_i)\\
&&
-\sum_{i=1}^d \log \Gamma(\alpha p_i-(1-\alpha)q_i)-
\\ && \alpha\log \Gamma(\sum_{i=1}^d  |p_i|)-(1-\alpha) \log \Gamma(\sum_{i=1}^d  |q_i|).
\end{eqnarray*}

Dirichlet distributions are exponential families of order $d$ with natural parameters $\theta=(p_1-1, ..., p_d-1)$ and log-normalizer
$F(\theta)=\sum_{i=1}^d \log \Gamma(\theta_i+1) -\log \Gamma(d+\sum_{i=1}^d \theta_i)$ 
(or $F(p)=\sum_{i=1}^d \log \Gamma(p_i) -\log \Gamma(\sum_{i=1}^d p_i)$).
Our work extends the computation of Chernoff $\alpha$-divergences to {\it arbitrary} exponential families using the natural parameters and the log-normalizer.

Since Chernoff information is defined as the {\it maximal} Chernoff $\alpha$-divergence (which corresponds to minimize the Chernoff coefficient in the Bayes error upper bound, with $0<c_\alpha(p,q)\leq 1$), we concentrate on  maximizing the equivalent skew Jensen divergence.

\section{Maximizing $\alpha$-Jensen divergences}\label{sec:JensenBregman}

We now prove that the maximal skew Jensen divergence can be computed as an {\it equivalent} Bregman divergence.
First, consider univariate functions.
Let $\alpha^*=\arg\max_{0<\alpha<1} J_F^{(\alpha)}(p:q)$ be the maximal $\alpha$-divergence.
Following Figure~\ref{fig:max}, we observe that we have {\it geometrically} the following relationships~\cite{entdivmean-1995}: 

\begin{equation}\label{eq:thales}
J_F^{(\alpha^*)}(p : q) = B_F(p : m_{\alpha^*}) = B_F(q : m_{\alpha^*}),
\end{equation}
where $m_\alpha =  \alpha p+(1-\alpha)q$ be the $\alpha$-mixing of distributions $p$ and $q$.
We maximize the $\alpha$-Jensen divergence by setting its derivative to zero:
\begin{equation}
\frac{\mathrm{d}J_F^{(\alpha)}(p:q)}{\mathrm{d}\alpha} = F(p)-F(q) - (m_\alpha)'   F'(m_\alpha).
\end{equation}

\begin{figure}
\includegraphics[width=0.9\columnwidth]{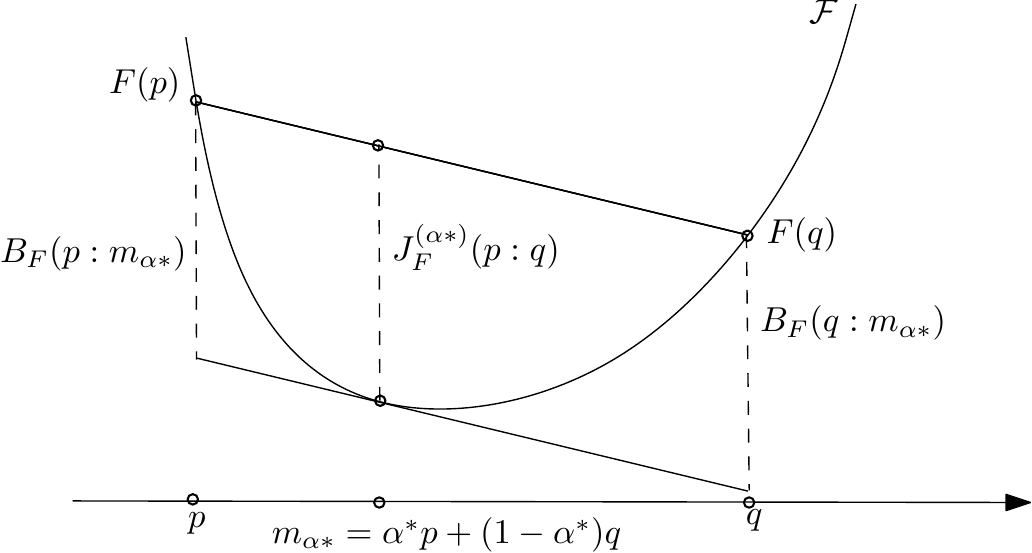}
\caption{The maximal Jensen $\alpha$-divergence is a Bregman divergence in disguise:
$J_F^{(\alpha^*)}(p:q) = \max_{\alpha\in(0,1)} J_F^{(\alpha)}(p:q)  = B_F(p : m_{\alpha^*}) = B_F(q : m_{\alpha^*})$.}
\label{fig:max}
\end{figure}

Since the derivative $(m_\alpha)'$ of $m_\alpha$ is equal to $p-q$, we deduce from the maximization that 
$\frac{\mathrm{d}J_F^{(\alpha)}(p:q)}{\mathrm{d}\alpha}=0$ implies the following constraint:

\begin{equation}
F'(m_\alpha^*)= \frac{F(p)-F(q)}{p-q}.  \label{eq:solvealpha}
\end{equation}

\def\alphastar{\frac{{F'}^{-1}\left(\frac{F(p)-F(q)}{p-q}\right)-p}{q-p}}
\def\betastar{\frac{q-F^{-1}\left(\frac{F(p)-F(q)}{p-q}\right)}{q-p}}

This  means geometrically that the tangent at $\alpha^*$ should be parallel to the line passing through $(p,z=F(p))$ and $(q,z=F(q))$, as illustrated in Figure~\ref{fig:max}.
It follows that 

\begin{equation}
\alpha^*=\alphastar.
\end{equation}

Using Eq.~\ref{eq:thales} , we have $p-m_\alpha^*=(1-\alpha^*)(p-q)$, so that it comes

\begin{eqnarray*}
B_F(p : m_\alpha^*) &=& F(p) - F(m_\alpha^*) - (1-\alpha^*)(F(p)-F(q))\\
&=& \alpha^* F(p) + (1-\alpha^*) F(q) -F(m_{\alpha^*})\\
&=& J_F^{(\alpha^*)}(p:q)
\end{eqnarray*}

Similarly, we have 
$q-m_\alpha^*=\alpha^*(q-p)$ and it follows that

\begin{eqnarray}
B_F(q : m_\alpha^*) &=& F(q) - F(m_\alpha^*)  - (q-m_\alpha^*) F'(m_\alpha^*) \nonumber\\
 & = & F(q) - F(m_\alpha^*) + \alpha^* (p-q) \frac{F(p)-F(q)}{p-q}  \nonumber\\
 & = & \alpha^* F(p) + (1-\alpha^*) F(q) - F(m_{\alpha^*}) \nonumber\\
 & = & J_F^{(\alpha^*)}(p:q)
\end{eqnarray}

Thus,  we {\it analytically} checked the geometric intuition  that $J_F^{(\alpha^*)}(p:q)=B_F(p : m_\alpha^*)=B_F(q : m_\alpha^*)$.
Observe that in the definition of a Bregman divergence, we require to compute explicitly the gradient $\nabla F$, but that in the Jensen  $\alpha$-divergence, we do not need it. (However, the gradient computation occurs in the computation of the best $\alpha$). 


\subsection{Single-parametric exponential families}\label{sec:order1}

We conclude that the Chernoff information divergence of members of the same exponential family of order $1$ has always a closed-form analytic formula:

\begin{eqnarray}
\lefteqn{C(p:q)=}\\ \label{eq:cf}
&&\alpha^*F(p) + (1-\alpha^*) F(q) - F\left(F'^{-1}(\frac{F(p)-F(q)}{p-q})\right), \nonumber
\end{eqnarray}
with 
\begin{equation}\label{eq:alphastar}
\alpha^*=\alphastar.
\end{equation}

\begin{figure}
\centering

\includegraphics[bb= 0 0 858 476,width=0.7\columnwidth]{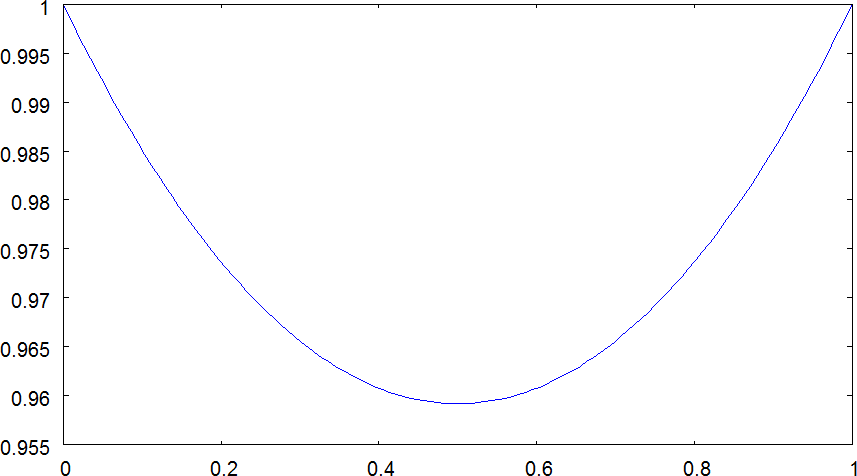}\\
\includegraphics[bb= 0 0 858 476,width=0.7\columnwidth]{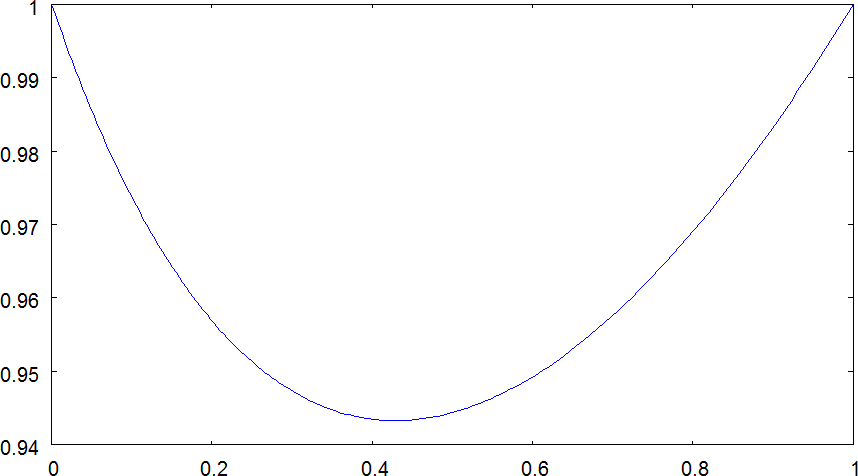}

\caption{Plot of the $\alpha$-divergences  for two normal distributions for $\alpha\in(0,1)$: (Top) $p\sim N(0,9)$ and $q\sim N(2,9)$, and (Bottom)
 $p\sim N(0,9)$ and $q\sim N(2,36)$. Observe that for equal variance, the minimum $\alpha$ divergence is obtained for $\alpha=\frac{1}{2}$, and that Chernoff divergence reduces to the Bhattacharyya divergence.  }
\end{figure}


Common exponential families of order $1$ include the Binomial, Bernoulli, Laplacian (exponential), Rayleigh, Poisson, Gaussian with fixed standard deviations. To illustrate the calculation method, let us instantiate the  univariate Gaussian and  Poisson   distributions.

For univariate Gaussian differing in mean only (ie., constant standard deviation $\sigma$), we have the following:

$$
\theta=\frac{\mu}{\sigma^2}, \qquad F(\theta)=\frac{\theta^2\sigma^2}{2}=\frac{\mu^2}{2\sigma^2}, \qquad
F'(\theta)=\theta\sigma^2=\mu
$$

We solve for $\alpha^*$ using Eq.~\ref{eq:alphastar}:

\begin{eqnarray*}
F'(\alpha^* \theta_p+(1-\alpha^*)\theta_q) & =& \frac{F(\theta_p)-F(\theta_q)}{\theta_p-\theta_q}\\
\mu_p+(1-\alpha^*)(\mu_q-\mu_p)  & = & \frac{\mu_p^2-\mu_q^2}{2(\mu_p-\mu_q)} = \frac{\mu_p+\mu_q}{2}
\end{eqnarray*}
It follows that $\alpha^*=\frac{1}{2}$ as expected, and that the Chernoff information is the Bhattacharrya distance:

\begin{eqnarray*}
C(p:q)& =& C_{\frac{1}{2}}(p,q)=J_{F}^{(\frac{1}{2})}(\theta_p:\theta_q),\\
& = & \frac{1}{2\sigma^2}(\frac{\mu_p^2+\mu_q^2}{2})-\frac{(\frac{\mu_p+\mu_q}{2})^2}{2\sigma^2}\\
& = & \frac{1}{8\sigma^2} (\mu_p-\mu_q)^2
\end{eqnarray*}

\def\lambdar{\frac{\lambda_2}{\lambda_1}}
For Poisson distributions ($F(\theta)=\exp(\theta)=F(\log\lambda)=\exp\log\lambda=\lambda$), Chernoff divergence is found by
first computing 

\begin{equation}
\alpha^*=\frac{\log\frac{\lambdar-1}{\log\lambdar}}{\log \lambdar}. \label{eq:poissonstar}
\end{equation}
Then using Eq.~\ref{eq:cf}, we deduce that

\begin{eqnarray}
C(\lambda_1:\lambda_2) & = & \lambda_2+\alpha^*(\lambda_1-\lambda_2) - \exp(m_{\alpha^*}) \nonumber \\
 &= & \lambda_2+\alpha^*(\lambda_1-\lambda_2) -\nonumber\\
 && \exp(\alpha^*(\log\lambda_1) + (1-\alpha^*)\log\lambda_2)\nonumber \\
 & = &  \lambda_2+\alpha^*(\lambda_1-\lambda_2) - {\lambda_1}^{\alpha^*}{\lambda_2}^{1-\alpha^*} \label{eq:poisson}
\end{eqnarray}

Plugging Eq.~\ref{eq:poissonstar} in Eq.~\ref{eq:poisson}, and ``beautifying'' the formula yields the following closed-form solution for the Chernoff information:

\begin{equation}
C(\lambda_1,\lambda_2)=\lambda_1  \frac{(\lambdar-1) (\log\frac{\lambdar-1}{\log\lambdar} -1) + \log\lambdar}{\log \lambdar}.
\end{equation}

\subsection{Arbitrary exponential families}\label{sec:multiorder}
For {\it multivariate} generators $F$, we consider the restricted univariate convex function $F_{pq}(\alpha)=F(p+(1-\alpha)(q-p))$ with parameters $p'=0$ and $q'=1$, so that $F_{pq}(0)=F(p)$  and $F_{pq}(1)=F(q)$. 
We have 
\begin{equation}
C_F(p:q) = \max_{\alpha} J_F^{(\alpha)}(\theta_p:\theta_q) = \max_{\alpha} J_{F_{\theta_p\theta_q}}^{(\alpha)}(0:1).
\end{equation}

We have $F'_{pq}(\alpha)= (p-q)^T \nabla F(\alpha p+(1-\alpha)q)$.
To get the inverse of $F'_{pq}$, we need to solve the equation:

\begin{equation}\label{solvealpha}
(p-q)^T \nabla F(\alpha^* p + (1-\alpha^* )q) = F(q)-F(p).
\end{equation}
Observe that in 1D, this equation matches  Eq.~\ref{eq:solvealpha}.
Finding $\alpha^*$ may not always be in closed-form. 
Let $\theta^*= \alpha^* p + (1-\alpha^* )q$, then we need to find $\alpha^*$ such that

\begin{equation}\label{eq:opt}
(p-q)^T \nabla F(\theta^*) = F(q)-F(p).
\end{equation}

Now, observe that equation~\ref{eq:opt} is equivalent to the following condition: 

\begin{equation}\label{eq:bisector}
B_F(\theta_p : \theta^*) = B_F(\theta_q : \theta^*)
\end{equation}
 and that therefore it follows that
 \begin{equation}
 \KL(p_F(x;\theta^*) : p_F(x;\theta_p)) =  \KL(p_F(x;\theta^*) : p_F(x;\theta_q)).
 \end{equation}

Thus it can be checked that the Chernoff distribution $r^*=p_F(x;\theta^*)$ is written as

\begin{equation}\label{eq:chernoffpt}
p_F(x;\theta^* ) =  \frac{p_F(x;\theta_p)^{\alpha^*}(x) p_F(x;\theta_q)^{1-\alpha^*}}{\int_x p_F(x;\theta_p)^{\alpha^*}(x) p_F(x;\theta_q)^{1-\alpha^*} \dx}
\end{equation}

\section{The Chernoff point}\label{sec:chernoffpoint}

Let us consider now the exponential family 
\begin{equation}
E_F=\{p_F(x;\theta)\ |\ \theta\in\Theta\},
\end{equation}
as a smooth statistical manifold~\cite{informationgeometry-2000}.
Two distributions $p=p_F(x;\theta_p)$ and $q=p_F(x;\theta_q)$ are geometrically viewed as two points (expressed as $\theta_p$ and $\theta_q$ coordinates in the natural coordinate system).
The Kullback-Leibler divergence between $p$ and $q$ is equivalent to a Bregman divergence on the natural parameters:
$\KL(p:q)=B_F(\theta_q:\theta_p)$. For infinitesimal close distributions $p\simeq q$, the Fisher information provides the underlying Riemannian metric, and is equal to the Hessian  $\nabla^2 F(\theta)$ of the log-normalizer for exponential families~\cite{informationgeometry-2000}.  
On statistical manifolds~\cite{informationgeometry-2000}, we define {\it two types} of geodesics: the mixture $\nabla^{(m)}$ geodesic and the exponential $\nabla^{(e)}$ geodesics:

\begin{eqnarray}
\nabla^{(m)}(p(x),q(x),\lambda) & = & (1-\lambda) p(x) + \lambda q(x),\\
\nabla^{(e)}(p(x),q(x),\lambda) & = & \frac{p(x)^{1-\lambda}  q(x)^{\lambda}}{\int_x p(x)^{1-\lambda}  q(x)^{\lambda} \dx  } ,\\
\end{eqnarray}

Furthermore, to any convex function $F$, we can associate a dual convex conjugate $F^*$ (such that ${F^*}^*=F$) via the Legendre-Fenchel transformation:
\begin{equation}
F^*(y)=\max_{x} \{ \innerproduct{x}{y}-F(x)\}.
\end{equation}
The maximum is obtained for $y=\nabla F(x)$. Moreover, the convex conjugates are coupled by reciprocal inverse gradient: $\nabla F^*=(\nabla F)^{-1}$. 
Thus a member $p$ of the exponential family, can be parameterized by its natural coordinates $\theta_p=\theta(p)$, or dually by its expectation coordinates $\eta_p=\eta(p)=\nabla F(\theta)$. 
That is, there exists a {\it dual coordinate system} on the information manifold $E_F$ of the exponential family.

Note that the Chernoff distribution $r^*=p_F(x;\theta^*)$ of Eq.~\ref{eq:chernoffpt} is a distribution belonging to the exponential geodesic.
The natural parameters on the exponential geodesic are interpolated linearly in the $\theta$-coordinate system.
Thus the exponential geodesic segment has natural coordinates $\theta(p,q,\lambda)=(1-\lambda)\theta_p+\lambda\theta_q$.
Using the dual expectation parameterization $\eta^*=\nabla F(\theta^*)$, we may also rewrite the optimality criterion of equation Eq.~\ref{eq:opt} equivalently as

\begin{equation}
(p-q)^T  \eta^* = F(q)-F(p),
\end{equation}
with $\eta^*$ a point on  the exponential geodesic parameterized by the expectation parameters (each mixture/exponential geodesic can be parameterized in each natural/expectation coordinate systems).

From Eq.~\ref{eq:bisector}, we deduce that the Chernoff distribution should {\it also} necessarily belong
to the right-sided Bregman Voronoi bisector 
\begin{equation}
V(p,q)=\{ x\ |\ B_F(\theta_p : \theta_x) = B_F(\theta_q : \theta_x)\}.
\end{equation}
This bisector is curved in the natural coordinate system, but affine in the dual expectation coordinate system~\cite{bvd-2010}.
Moreover, we have $B_{F}(q:p)=B_{F^*}(\nabla F(p):\nabla F(q))$, so that we  may express the right-sided bisector equivalently in the expectation coordinate system as 
\begin{equation}
V(p,q)=\{ x\ |\ B_{F^*}(\eta_x : \eta_p) = B_F(\eta_x: \eta_q)\}.
\end{equation}
 That is, a left-sided bisector for the dual Legendre convex conjugate $F^*$.

\begin{figure}
\centering
\includegraphics[width=0.8\columnwidth]{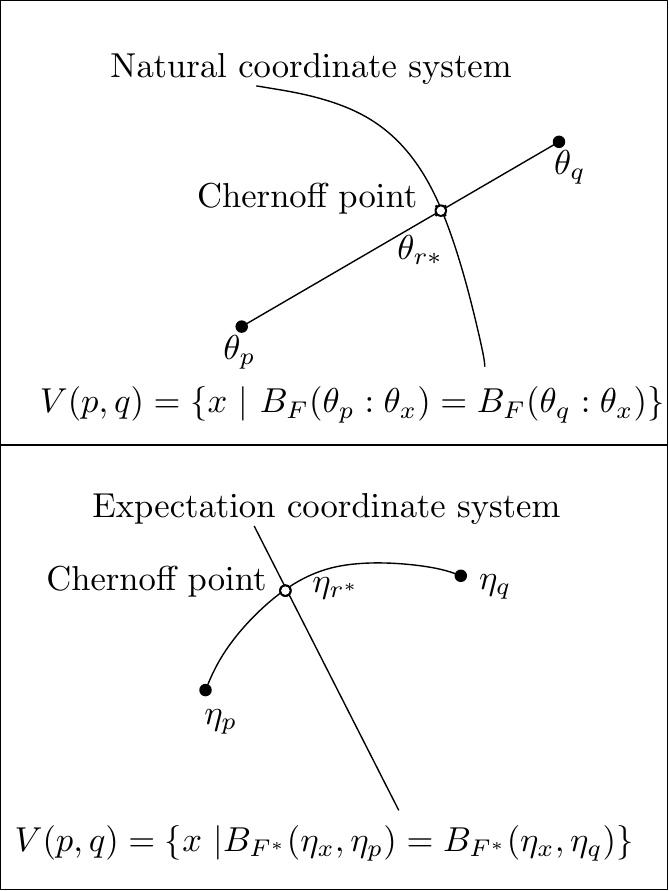}
\caption{Chernoff point $r^*$ of $p$ and $q$ is defined as the intersection of the exponential geodesic $\nabla^{(e)}(p,q)$ with the right-sided Voronoi bisector $V(p,q)$. In the natural coordinate system, the exponential geodesic is a line segment and the right-sided bisector  is curved.
In the dual expectation coordinate system, the exponential geodesic is curved, and the right-sided bisector is affine.}\label{fig:chernoffpoint}
\end{figure}

Thus the Chernoff distribution $r^*$ is viewed as a {\it Chernoff point} on the statistical manifold such that $r^*$ is defined as
the intersection of the exponential geodesic ($\eta$-geodesic, or $e$-geodesic) with the curved bisector $\{ x\ |\ B_F(\theta_p : \theta_x) = B_F(\theta_q : \theta_x)\}$.
In~\cite{bvd-2010}, it is proved that the exponential geodesic right-sided bisector intersection is Bregman orthogonal.
Figure~\ref{fig:chernoffpoint} illustrates the geometric property of the Chernoff distribution (which can be viewed indifferently in the natural/expectation parameter space), from which the corresponding best exponent can be retrieved to define the Chernoff information.

We following section builds on this {\it exact geometric characterization} to build a geodesic bisection optimization method to arbitrarily finely approximate the optimal exponent.

%
%
%
%
%
%

\section{A geodesic bisection algorithm}
\label{sec:primaldual}

To find the Chernoff point $r^*$ (ie., the parameter $\theta^*=(1-\alpha^*)\theta_p+\alpha^*\theta_q$, a simple bisection algorithm follows: 
Let initially $\alpha\in [\alpha_{m},\alpha_M]$ with $\alpha_{m}=0,\alpha_M=1$. 
Compute the midpoint $\alpha'=\frac{\alpha_m+\alpha_M}{2}$ and let $\theta=\theta_p+\alpha'(\theta_q-\theta_p)$.
 If $B_F(\theta_p:\theta)<B_F(\theta_q:\theta)$ recurse on interval $[\alpha',\alpha_M]$, otherwise recurse on interval $[\alpha_m,\alpha']$.
At each stage we split the $\alpha$-range in the $\theta$-coordinate system.
Thus we can get arbitrarily precise approximation of the Chernoff information of members of the same exponential family by walking on the exponential geodesic towards the Chernoff point.

\section{Concluding remarks}\label{sec:Conclusion}

Chernoff divergence upper bounds asymptotically the optimal Bayes error~\cite{ct-1991}: $\lim_{n\to\infty} E^* = e^{-n C(p : q)}$.
Chernoff bound thus provides the best Bayesian exponent error~\cite{ct-1991}, improving over the Bhattacharyya divergence ($\alpha=\frac{1}{2}$):
\begin{equation}
\lim_{n\to\infty} E^* = e^{-n C(p,q)}\leq e^{-n B(p,q)},
\end{equation}
 at the expense of solving an optimization problem.
The probability of misclassification error can also be lower bounded by information-theoretic statistical distances~\cite{tightBayes-1996,skl-2001} (Stein lemma~\cite{ct-1991}):
\begin{equation}
\lim_{n\to\infty} E^* = e^{-n C(p : q)}  \geq e^{-n R(p : q)}  \geq e^{-n J(p : q)},
\end{equation} 
where $J(p:q)$ denotes half of the Jeffreys divergence $J(p:q)=\frac{\KL(p:q)+\KL(q:p)}{2}$ (i.e., the arithmetic mean on sided relative entropies) and $R(p:q)=\frac{1}{\frac{1}{\KL(p:q)}+\frac{1}{\KL(p:q)}}$ is the resistor-average distance~\cite{skl-2001} (i.e., the harmonic mean).
In this paper, we have shown that the Chernoff $\alpha$-divergence of members of the same exponential family can be computed from an equivalent $\alpha$-Jensen divergence on corresponding natural parameters. 
Then we have explained  how the maximum $\alpha$-Jensen divergence yields a simple gradient constraint. 
As a byproduct this shows that the maximal $\alpha$-Jensen divergence is equivalent to compute a Bregman divergence.
For  single-parametric exponential families (order-$1$ families or dimension-wise separable families), we deduced a closed form formula for the Chernoff  divergence (or Chernoff information).
Otherwise, based on the framework of information geometry, we interpreted the optimization task as of finding the ``Chernoff point'' defined by the intersection of the exponential geodesic linking the source distributions with a right-sided Bregman Voronoi bisector.
Based on this observation, we designed an efficient geodesic bisection algorithm to arbitrarily approximate the Chernoff information.

\begin{IEEEbiographynophoto}{Frank Nielsen}
defended his PhD thesis on Adaptive
Computational Geometry in 1996 (INRIA/University of Sophia-Antipolis, France), and his accreditation to lead research in 2006. 
He is a researcher of Sony Computer Science
Laboratories Inc., Tokyo (Japan) since 1997, and a professor at \'Ecole Polytechnique since 2008.
His  research focuses on computational information geometry with applications to visual computing.
\end{IEEEbiographynophoto}

\end{document}